\documentclass[%
 aip,
apl,%
 amsmath,amssymb,
 reprint,%
]{revtex4-1}

\usepackage{graphicx}
\usepackage{dcolumn}
\usepackage{bm}
\usepackage{subfig}

\begin{document}

\preprint{AIP/123-QED}

\title{Soft-proton exchange on Magnesium-oxide-doped substrates: a route toward efficient and power-resistant nonlinear converters.}
\affiliation{Universit\'e C\^ote d’Azur, CNRS, Institut de Physique de Nice, France
}
\author{T. Lunghi}
\email{tlunghi@unice.fr}

\author{F. Doutre}
\author{G. Legoff}
\author{G. Ayenew}
\author{H. Tronche}
\author{S. Tanzilli}
\author{P. Baldi}
\author{M. De Micheli}


\date{\today}

\begin{abstract}
Despite its attractive features, Congruent-melted Lithium Niobate (CLN) suffers from Photo-Refractive Damage (PRD). This light-induced refractive-index change hampers the use of CLN when high-power densities are in play, a typical regime in integrated optics. The resistance to PRD can be largely improved by doping the lithium-niobate substrates with magnesium oxide. However, the fabrication of waveguides on MgO-doped substrates is not as effective as for CLN: either the resistance to PRD is strongly reduced by the waveguide fabrication process (as it happens in Ti-indiffused waveguides) or the nonlinear conversion efficiency is lowered (as it occurs in annealed-proton exchange). Here we fabricate, for the first time, waveguides starting from MgO-doped substrates using the Soft-Proton Exchange (SPE) technique and we show that this third way represents a promising alternative. We demonstrate that SPE allows to produce refractive-index profiles almost identical to those produced on CLN without reducing the nonlinearity in the substrate. We also prove that the SPE does not affect substantially the resistance to PRD. Since the fabrication recipe is identical between CLN and MgO-doped substrates, we believe that SPE might outperform standard techniques to fabricate robust and efficient waveguides for high-intensity-beam confinement.
\end{abstract}
 
\pacs{Valid PACS appear here}
\maketitle

\section{Introduction}
Wavelength nonlinear conversion is an attractive approach to obtain coherent radiation in regions of the spectrum where lasers are unavailable or impractical\cite{Richter1998, Richter2005}. For example, up-conversion of photons from telecom to near-infrared  wavelengths would allow to benefit of the rapidity, compactness and tunability of the telecommunication laser sources in a spectral regime where comparable sources are missing\cite{Lutfi2015}. Among the different nonlinear materials, congruent Lithium Niobate (CLN) is one of the most widely used, because it features one of the largest nonlinear coefficient, a reliable crystal-growth process, and wide transparency range (between 350\,nm and 5200\,nm). Reliable techniques have been developed to push further its characteristics: through quasi-phase-matching\cite{Armstrong1962} the momentum-conservation relations can be engineered allowing to address a wide set of three-wave mixing processes. In addition several techniques have been set up to integrate waveguides inside the substrate. Among these techniques, the arguably easiest, fastest, cheapest, and most flexible way consists in replacing lithium ions with protons to locally increase the refractive index\cite{Jackel1982} (\emph{proton exchange}). Technically, the CLN specimens are immersed into a hot liquid source of protons. The exchange temperature plays a fundamental role\cite{DeMicheli1986}: at moderate temperatures (between 150\,$^{\circ}$C to 200\,$^{\circ}$C) proton diffusion inside the crystal is extremely slow and protons tend to accumulate at the surface. This alters severely the crystalline structure and forces several subsequent annealing steps to partially recover the original nonlinearity (\emph{Annealed Proton Exchange, APE}\cite{Bortz1991}). Alternatively, a proton exchange at higher temperatures (300$^{\circ}$C-350$^{\circ}$C) increases proton diffusion speed in the substrate and preserves the crystallographic structure\cite{Chanvillard2000} together with most of the material properties (\emph{Soft Proton Exchange, SPE}). Although this technique results in higher nonlinear conversion efficiency, it is rarely employed since it requires a more delicate fabrication technique.%

Despite its numerous advantages, CLN suffers from photorefractive damage (PRD)\cite{Glass1975}. Since the light travelling  through the crystal gets absorbed, an excess of free electrons is photo-generated in the illuminated area. These charges are free to move away from the illuminated region but they gets trapped just outside this area where the photoconductivity is rather suppressed. There they form a space-charge field which interacts with the electro-optic properties of the crystal: the refractive-index profile is altered inducing a modification of the light distribution which modifies again the space-charge field. This results in a fluctuation of the intensity profile that becomes more important as the power densities increase\cite{Nava13}. %
Mitigation approaches to de-trap photogenerated electrons include increasing operating temperature, modifying the crystallographic structure (stoichiometric LN\cite{Jermann1995}) or doping the materials with ions, as it happens for magnesium-oxide-doped lithium niobate (MgO-doped LN)\cite{Bryan1984}. Due to MgO-doped LN’s commercial availability and PRD resistance, this solution is ideal, however the fabrication of waveguide using APE technique is challenging: during the exchange the crystallographic structures is altered permanently\cite{Sun2015, Korkishko2003} and even if the nonlinear coefficient might be partially recovered, MgO-doped LN waveguides feature a lower conversion efficiency compared to CLN\cite{Roussev2006}. This pushes the researchers to investigate other dopants\cite{Langrock2016}. It is also known that the APE lowers the resistance to PRD\cite{Steinberg95} reducing the advantage of using MgO-doped substrates. 

In this paper, we report about the fabrication of SPE waveguides on MgO-doped LN substrates. We show that this technique allows to produce similar refractive index profiles in both MgO-doped LN and CLN. We also verify that the nonlinear coefficient is preserved confirming previously proposed hypothesis\cite{Korkishko2003}. From the practical point of view, the fabrication technique is substantially identical to the one used to fabricate waveguide on standard CLN and therefore can be immediately transferred to the MgO-doped LN substrates. The qualitative tests performed on these waveguides show that the PRD resistance is, at least partially, preserved even after the exchange. These results prove that SPE might outperform standard APE to produce robust nonlinear waveguides for high-intensity-beam confinement.
\newline

\section{Sample fabrication} %
To accurately determine the proton dose necessary to realize SPE, several slab waveguides were fabricated starting from optically graded \mbox{z-cut} LN substrates containing 5 mol\% MgO, supplied by the Fujian Institute of Research. The exchange is performed in a glass ampoule\citep{Chanvillard2000} with a shrinkage in its middle point. The ampoule is initially filled with a powder mixture of acid benzoic~(AB) buffered with a small quantity of lithium benzoate~(LB). The percentage of buffer compound in the mixture \mbox{($\rm{\rho_{LB}}$)} determines the concentration of protons available during the reaction. However small concentrations of water in the mixture increase drastically the proton dose during the exchange and degrade the reproducibility of the fabrication process\cite{Mushinsky2015}. Therefore we prepare the mixture and fill the ampoule in an dry environment (lower than \mbox{5\%rh}). The sample under test is then inserted in the ampoule and, thanks to the shrinkage in the middle, it is initially separated from the powder. To further reduce the amount of residual water in the mixture the pressure in the glass ampoule is lowered below \mbox{3$\times$10$^{-6}$ bar}. Finally the ampoule is hermetically sealed and placed in an oven at \mbox{300~$^{\circ}$C}. After one hour later, the temperature in the oven is uniform and the acid mixture is melted. The ampoule is flipped upside down to start the exchange. At the end of the exchange, the ampoule is flipped back and removed from the oven. 
\newline

\section{Refractive-index profile} 
\begin{figure}[!htb]
\begin{center}
\includegraphics[width=0.9\columnwidth]{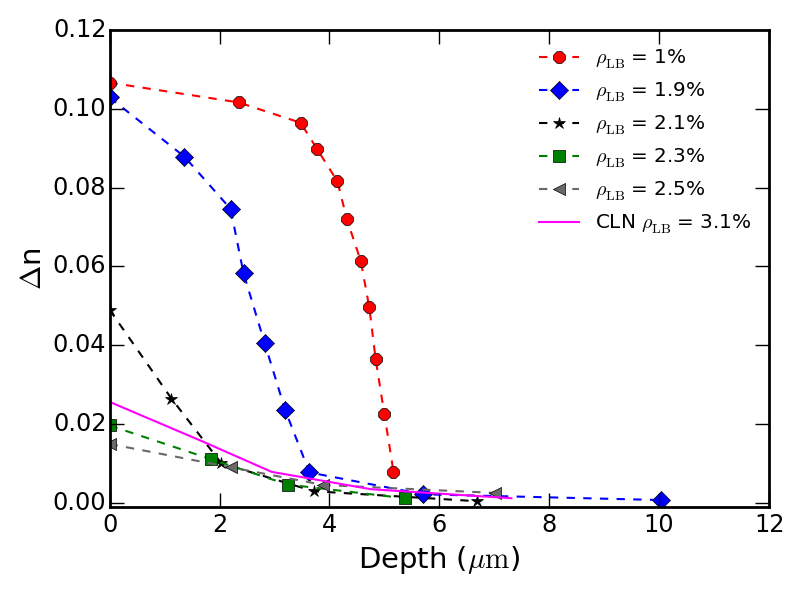}
\end{center}
\caption{Refractive-index profiles reconstructed using the Inverse-WKB technique for different $\rho_{\rm{LB}}$. The exchange duration corresponds to 24 hours for sample $\rho_{\rm{LB}}$~=~1\% (red dotted line, circle markers) and 72~hours otherwise. For comparison, a typical profile for a SPE waveguide fabricated on a CLN substrate is added (magenta solid line). The refractive index of the substrate is subtracted for clarity.}
\label{fig:Mlines}
\end{figure}

After the exchange, the refractive-index profiles of each sample have been characterized. We initially fabricate several slab waveguides with \mbox{$\rm{\rho_{LB}}$} varying between 1\% to 2.7\%. All the exchanged samples are multimode at the characterization wavelength of 633~nm. The effective index of each guided mode is assessed by measuring the excitation angles of bright m-lines with a two-prism-coupler setup\cite{Tien70}. These effective indices allow to reconstruct the refractive-index profile using the inverse-WKB (iWKB) method\cite{White1976, Dikaev1981}. Figure~\ref{fig:Mlines} shows the measured profiles for different \mbox{$\rm{\rho_{LB}}$}. The refractive index of the substrate (n$_{sub} = 2.19175$) is subtracted. The exchange duration corresponds to 24~hours for the sample with \mbox{$\rm{\rho_{LB}}$~=~1\%} and 72~hours otherwise.%
Three different regimes can be identified. At \mbox{$\rho_{\rm{LB}} \approx$ 1\%} the refractive-index profile assumes a step-like shape: due to the strong acidity of the bath too many protons penetrate in the crystal. The increase of refractive index, \mbox{$\Delta$n}, is not proportional to the proton concentration and saturates at 0.107. This behaviour is typical in proton-exchanged waveguides\cite{Chen1993,Chen1994}. When the bath acidity is decreased, \mbox{$\rho_{\rm{LB}} >$~2.1\%}, the proton-substitution rate at the surface is lowered: the refractive-index profile assumes an exponential shape with $\Delta$n decreasing with higher $\rho_{\rm{LB}}$. 
When $\rho_{\rm{LB}}$ is comprised between 2.1\% and 1\% the refractive index profile assumes both the shapes: in the superficial layer \mbox{$\Delta$n} saturates with an almost step-like profile; deeper in the substrate the exponential tail is still present. 
For comparisons, we add in Fig.~\ref{fig:Mlines} a typical $\Delta$n for a SPE waveguide based on CLN substrate and exchanged for 72~hours at \mbox{300~$^{\circ}$C} with $\rho_{\rm{LB}}$~=~3.1\%. This comparison shows that MgO-doped substrates require a lower $\rho_{\rm{LB}}$ than in CLN to attain similar $\Delta$n. This might be due to a smaller proton substitution rate in MgO-doped substrates than in CLN substrates.
\newline

\section{Nonlinear-coefficient characterization} 
In order to prove that the exchange preserves the nonlinearity the local value of $\rm{d_{33}}$ has been probed by means of a standard scanning-SHG-microscopy setup\cite{ElHadi1997, Denev2011, Ahlfeldt1994}: a strong optical beam from a femto-second laser at 1550~nm is focused on a polished end-facet of the sample. A piezoelectric base moves the sample vertically allowing to make a depth profiling of the nonlinear response and probe both the exchanged and the un-exchanged region. After reflection, the fundamental beam ($P_{\omega}$) and the second-harmonic beam ($P_{2\omega}$), proportional to $\rm{d_{33}^2}$, are collected-back and separated using a dichroic mirror. $P_{\omega}$ is used as a profile reference indicating the depth position, $P_{2\omega}$ is used to probe the nonlinear coefficient. 

\begin{figure}[!htbp]
\begin{center}
\includegraphics[width=0.9\columnwidth]{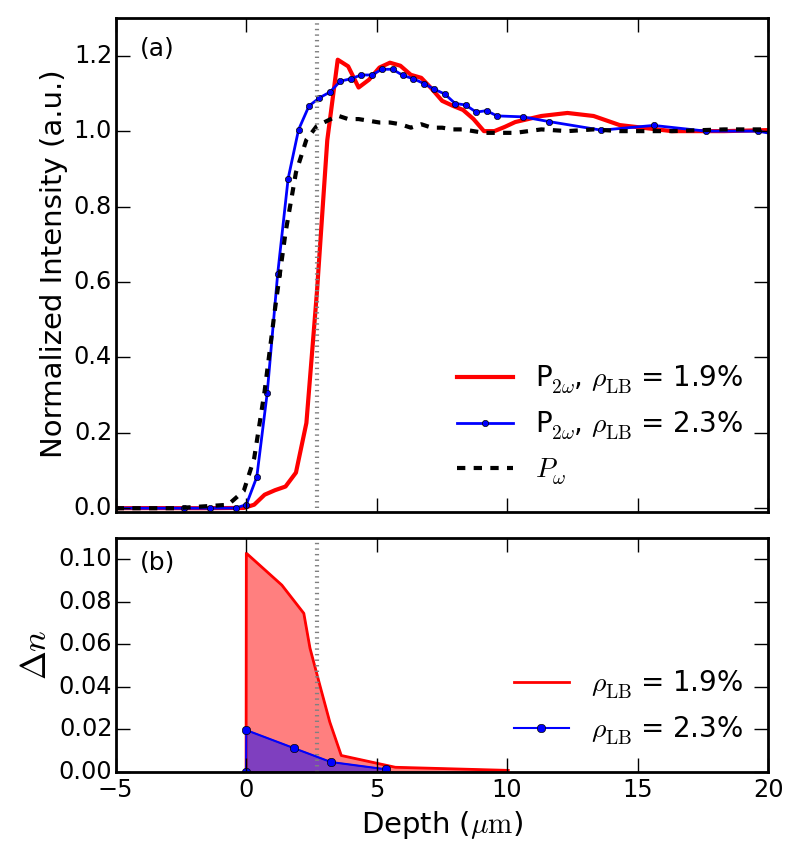}
\end{center}
\caption{(a) Reflected fundamental ($P_{\omega}$) and second harmonic intensities ($P_{2\omega}$) for two slab waveguides based on MgO-doped substrates. The backreflected fundamental signal is used as a profile reference indicating the depth position. (b) Refractive index profile at 633\,nm. The samples were fabricated with $\rho_{LB}$ equal to 1.9\% (red solid line) 2.3\% (blue line with dots). The latter sample shows an unaltered SHG signal indicating a good SPE.}
\label{fig:surfaceSHG}
\end{figure}

\mbox{Figure~\ref{fig:surfaceSHG}(a)} shows the SHG signal intensities as a function of the depth for two MgO-doped slab waveguides with \mbox{$\rm{\rho_{LB}}$= 2.3\,\%} (blue solid line with dot) and \mbox{$\rm{\rho_{LB}}$ = 1.9\,\%} (red solid line). The intensities are normalized relative to the signal well deep in the substrate which represents the value for the un-exchanged region. $P_{2\omega}\rm{(depth)}$ for sample \mbox{$\rm{\rho_{LB}}$= 2.3\,\%} spatially overlaps with $P_{\omega}$ and its intensity is equal or higher than the intensity in the un-exchanged region. This proves that the exchanged layer presents the same nonlinear coefficient as the substrate. In comparison the sample exchanged at \mbox{$\rm{\rho_{LB}}$ = 1.9\,\%} shows a different signature: $P_{2\omega}$ is equal to zero within the firsts 2-3~$\mu$m and it increases only deeper in the substrate. \mbox{Figure~\ref{fig:surfaceSHG}(b)} shows the \mbox{refractive-index} profiles for the two samples: the region where $P_{2\omega}$  is suppressed is compatible with the step region of the index profile of the waveguide. Note that $P_{2\omega}$ in the waveguide might be higher than the response obtained in the unexchanged substrate. This has been identified as an artefact associated with the guiding effect of the waveguide\citep{Oleksander2014} and should not be confused with an increase of the nonlinearities.
\section{Waveguide fabrication and photorefractive characterization}
\begin{figure}
	\centering
	\subfloat[]{
	  \includegraphics[width = 0.8\columnwidth]{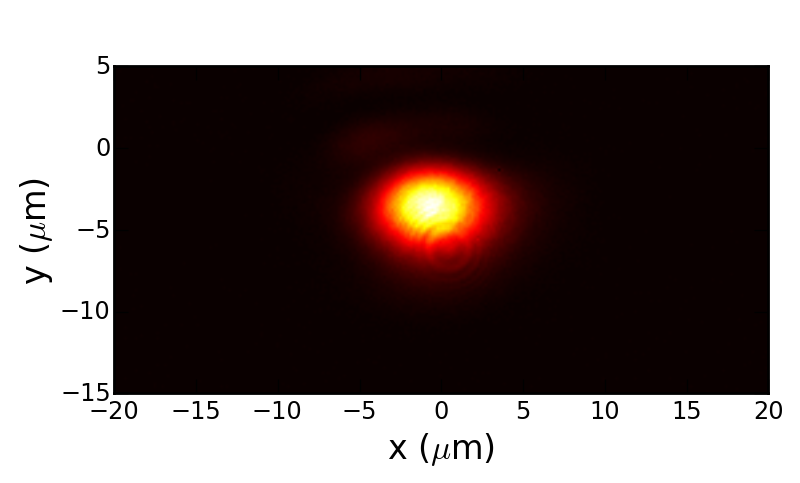}
	  \label{fig:beam}
	}
	\hspace{0mm}
	\subfloat[]{
	  \includegraphics[width = 0.5\columnwidth]{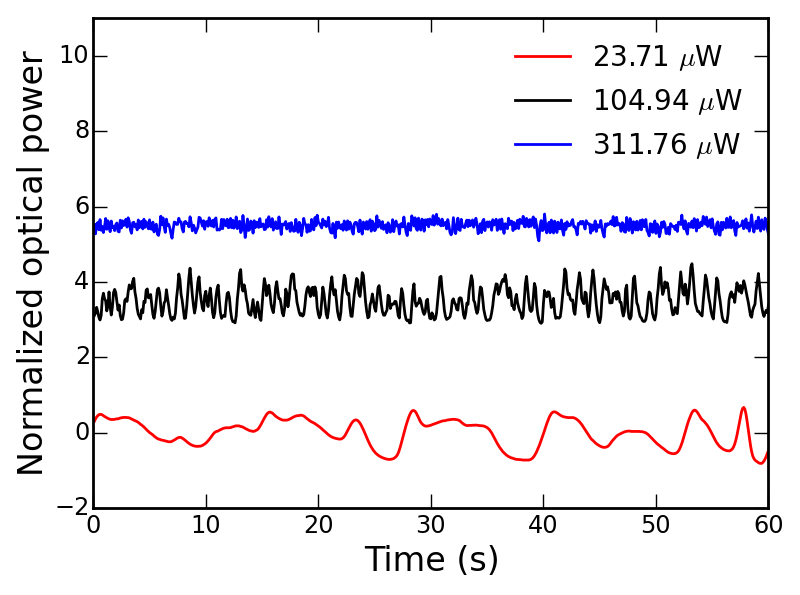}
	  \label{fig:photorefractiveCLN}
	}
	\subfloat[]{
	  \includegraphics[width= 0.5\columnwidth]{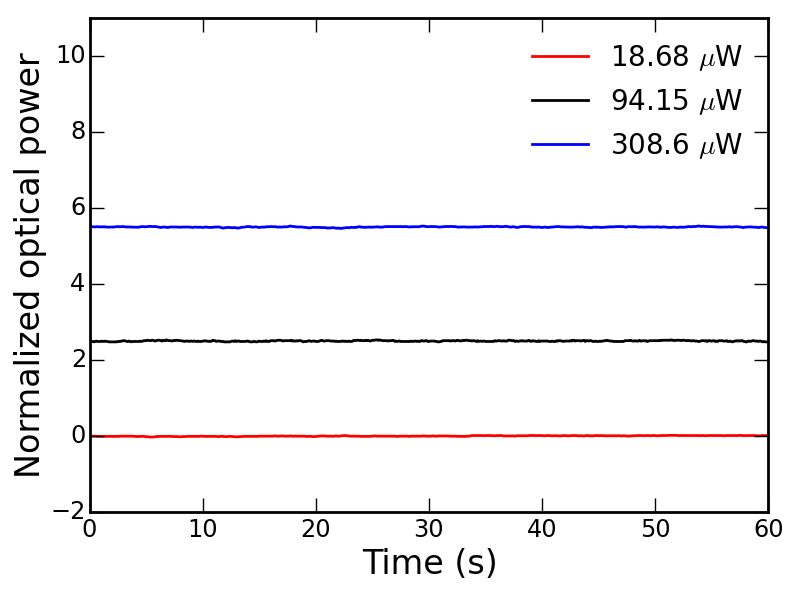}
	  \label{fig:photorefractiveMg}
	}
	\caption{(a) Measured intensity profile of the guided mode at 1550~nm. (b) Optical power transmitted trough a waveguide fabricated on a CLN substrate. The intensity is normalized with the average power for each trace. Traces are arbitrarily shifted for clarity. During the measurements, the power meter is able to collect only a portion of the beam so that beam fluctuations induced by the photorefractive damage are mapped in power fluctuations. Each trace corresponds to a different output power. The fluctuations are faster as the output power increases. This shows that the material suffers from PRD even at low power. (c) Same measurement for a photonic chip fabricated on a Mg-doped substrate. The measured optical power does not fluctuate since the optical beam is stable. This confirms that, at least with such a input power, the material does not suffers from PRD.}
\end{figure}

On the basis of the previous results, several photonic chips were exchanged starting from MgO-doped substrates. The fabrication started with the deposition of a photoresist layer on top of the -z face of the substrate. Then the desired pattern was transferred from a mask to the photoresist using standard lithographic techniques. A buffer layer of SiO$_2$ was deposited by means of RF sputtering. Through a lift-off technique, the photoresist left on the substrate was removed together with the silica above. These channels define where the proton exchange occurs while the layer of SiO$_2$ shields the substrate and prevents the proton substitution. The lithographic mask used during this test consists of straight waveguide channels with 6\,$\mu$m width. The exchange lasts 72 hours at 300$^{\circ}$C. 
Initial tests were conducted with a buffer layer 270\,nm thick and $\rho_{\rm{LB}}$ = \mbox{2.3\%}. Although these waveguides were single mode at 1550\,nm (see Fig.\ref{fig:beam}) and with low transmission losses (estimated smaller than 0.1 dB/cm), we identified the presence of a superficial planar waveguide everywhere below the SiO$_2$ mask. This is due to the permeability of the SiO$_2$ layer. While this waveguide was too shallow to guide light at telecom wavelengths, its presence should be avoided in order to increase the reproducibility of the process. Therefore, we increased the buffer SiO$_2$ layer up to 400\,nm-thick and we slightly increased $\rho_{\rm{LB}}$ to  \mbox{2.4\%}. After the exchange, samples are diced and optically polished in order to inject the light from the lateral facets.\\ 

Previous study has shown that APE waveguides in MgO-doped substrate are still  affected by the photorefractive damage\cite{Steinberg95}. Therefore we prepared two samples with similar index profiles, one in CLN and the other in MgO-doped LN to compare their behaviour for increasing injected power. 

The samples under test were placed on a waveguide holder: the input light was injected from one lateral facet using a 5\,mm lens. The polarization of the input beam is oriented using a bulk polarization controller. The light outcoming from the waveguide was collected using a \mbox{f = 11\,mm} lens and sent toward a visible power meter. 
%
Both the CLN and the MgO-doped sample were characterized at 30$^{\circ}$C by injecting in the straight waveguide a 710\,nm beam generated by a 76-MHz mode-locked laser with picosecond pulses. Short wavelength and pulsed regime were intentionally chosen to amplify the effect of the photorefractive damage. The transmitted light is firstly measured with a power meter for different input powers. Then the output beam is slightly defocalized so that only a portion of the beam is impinging in the power meter. Due to PRD, the mode in the waveguide becomes unstable as the input power increases. This mode fluctuations introduce power fluctuations that can be measured with the power meter. Figure~\ref{fig:photorefractiveCLN} shows the power fluctations, measured by the power meter for different input powers. The average value is shifted for clarity. The label for each trace refers to the total power coupled in the waveguide and measured at the output. For an input average power larger than 100\,$\mu$W we noticed that frequency of the fluctuations reaches the bandwidth of the power meter. This gives the impression that the fluctuations are smaller at higher input power. 
Figure~\ref{fig:photorefractiveMg} shows the same measurement for a waveguide fabricated starting from an Mg-doped substrate. The power fluctuations are suppressed due to higher resistance to the photorefractive damage. This proves that the resistance to PRD is higher than in CLN. Although this result is only qualitative, the beam characteristics, i.e. wavelength and input power, correspond to a typical settings for a quantum integrated experiments\cite{Alibart2016}. Therefore this result is promising to support further investigations.\newline

\section{Conclusions and Outlooks}
CLN is one the most used material for nonlinear-wavelength conversion. Despite its attractive features, it suffers from PRD: strong input power alters the refractive index of the material as well as the beam profile. The resistance to PRD can be drastically increased by doping the lithium niobate substrates with other materials, such as Mangnesium-oxide. However for this kind of doped substrates, the standard waveguide fabrication techniques are not so efficient than in CLN. In particular, the waveguides produced by APE, one of the most used techniques, present lower conversion efficiencies. Unlike APE, SPE is a viable alternative that offers high nonlinear efficiency and a relatively-easy fabrication process. In this paper we report the first realization of low loss SPE waveguides in  MgO-doped LN substrates. We show that, with a slightly modified recipe,  refractive-index profiles almost identical to those obtained in CLN can be produced. We also show that the exchange does not reduce the nonlinearity and does not affect significantly the resistance to the PRD. This work demonstrates that SPE provides clear advantages compared to other techniques to produce efficient and power-resistant nonlinear waveguide.

\section{acknowledgement}
This work was supported financially by the FP7 program of the
European Commission through the Marie Curie ITN Picque project (grant agreement No. 608062), the Agence Nationale de la Recherche through the INQCA project (grant No. PN-II-ID-JRP-RO-FR-2014-0013). We thank O.~Alibart for useful discussions and Prof. Liang Wanguo of the Fujian Institute of Research for providing us with the Mg-doped crystals.

\nocite{*}
\bibliography{biblio}
\end{document}